# Coupled VO$_2$ oscillators circuit as analog first layer filter in convolutional neural networks

**Elisabetta Corti[1], Joaquin Antonio Cornejo Jimenez[1], Kham M. Niang[2], John Robertson[2], Kirsten E. Moselund[1], Bernd Gotsmann[1], Adrian M. Ionescu[3] and Siegfried Karg[1]**

[1]IBM Research Zürich, Rüschlikon, Switzerland

[2]Department of Engineering, University of Cambridge, Cambridge, United Kingdom

[3]Nanoelectronic Devices Laboratory, École Polytechnique Fédérale de Lausanne, Lausanne, Switzerland

**\* Correspondence:**
Siegfried Karg
sfk@zurich.ibm.com

## Abstract

In this work we present an in-memory computing platform based on coupled VO$_2$ oscillators fabricated in a crossbar configuration on silicon. Compared to existing platforms, the crossbar configuration promises significant improvements in terms of area density and oscillation frequency. Further, the crossbar devices exhibit low variability and extended reliability, hence, enabling experiments on 4-coupled oscillator. We demonstrate the neuromorphic computing capabilities using the phase relation of the oscillators. As an application, we propose to replace digital filtering operation in a convolutional neural network with oscillating circuits. The concept is tested with a VGG13 architecture on the MNIST dataset, achieving performances of 95% in the recognition task.

## 1    Introduction

Convolutional Neural Networks (CNNs) are the architecture of choice to compute image recognition tasks. Widely used in commercial technology for their recognition accuracy, they are hindered in speed and power efficiency by the frequent access to the memory they need to perform to train a high number of parameters for each convolutional layer in deep networks (Sebastian et al., 2020). The development of neuro-inspired hardware holds the promise of accelerating these algorithms by exploiting in-memory computing concepts and limiting the number of accesses to the memory. A system of coupled oscillator, or Oscillatory Neural Network (ONN) can be used to store and recognize multiple patterns in compact networks. As described in (Hoppensteadt and Izhikevich, 1999; Izhikevich, 2000), systems of coupled oscillators lock in frequency and establish programmable phase relations that can be used for in-time computing applications. An ONN comprises a system of oscillators, in the role of neurons, connected to each other with synaptic weights, that represent the strength of the oscillators' coupling and the memory of the network. The ONN systems therefore rely on encoding and processing the information

**Coupled VO₂ oscillators circuit as analog first layer filter in convolutional neural networks**

with time-delays in the circuits, rather than the amplitude of a signal, therefore being resilient to voltage noise and to scaled power supply.

Exploiting the associative memory capabilities of such networks, tasks as image recognition can be performed. Numerous works have simulated through mathematical and circuit simulations the digit pattern retrieval with different coupled oscillators technologies: (Cells et al., 2015) shows 20-pixel digit recognition using transition metal oxides and resistive ram technology; (Nikonov et al., 2015; Liyanagedera et al., 2016), perform similar simulations on Spin Torque Oscillators (STOs); (Hölzel and Krischer, 2011) with Van der Pool oscillators. These works are based on storing patterns with $n \, x \, m$ pixels into an ONN that comprises $n \, x \, m$ oscillators. To perform the recognition, a distorted pattern of the same pixel size is fed to the ONN, and using the minimum phase attractor of the circuit, the right stored pattern is retrieved. The output is an $n \, x \, m$ pixels image of the correct digit. Image classifications tasks however work quite differently. Taking as an example digit classification through a neural network, an image of $n \, x \, m$ pixels if fed into the network. The network output is an $1 \, x \, 10$ array containing the classification probabilities of that image. This classification operation is most commonly performed by convolutional neural networks, that process the image with a series of trainable convolutional filters in the first few layers and achieve recognition after some fully-connected layers **(figure 1)**.

The link between convolutional neural networks and the associative memory capabilities of oscillatory neural networks has so far been explored, to our knowledge, only in (Liu and Mukhopadhyay, 2018), where an associative memory bank (Hopfield network) replaces the fully-connected layers in CNNs. The associative memory is here used to perform a classification with a combination of unsupervised learning and transfer learning techniques. Even though the concept is very interesting and promising, the Hopfield network that this technique envisions comprises between 256 and 2024 neurons. However, the physical demonstrations of oscillatory neural networks features maximum 100 oscillators as neurons with standard Phase Locked Loop or equivalent CMOS technology (Jackson et al., 2019). The technological challenge in the physical realization of large oscillatory neural networks resides in the complex dynamics of the oscillators' frequency and phase synchronization when the electrical components are affected by variability. This is even more true when the ONNs are built with novel oscillator technology, such as STOs or vanadium-dioxide (VO₂) oscillators (Romera et al., 2018; Raychowdhury et al., 2019), for which a maximum of 6 oscillators have been connected into a network.

Alternatively, it is suggested that the computing capabilities of small oscillator networks, with up to 10 nodes, can be efficiently exploited to do various image processing tasks, like graph coloring or image saliency processing (Cotter et al., 2014; Tsai et al., 2016). These demonstrations are however missing the link to show how ONNs can be used for image classification tasks. In this work we show the potentiality of the ONN technology for the realization of reconfigurable CNN in hardware, therefore bridging the gap between previous demonstration of ONN pattern retrieval and the industry-standard algorithms.

Among the new oscillator technologies, we concentrate our analysis on VO₂ oscillators, as they offer the advantage of realization of very compact oscillators, which can be easily coupled with standard electrical components, allowing for easy reconfigurability of the system (Parihar et al., 2015; Corti et al., 2018). VO₂ based oscillators also offer good scalability perspective and demonstrate operating voltage of less than 1 V and low power consumption (~ 20 µW per oscillator) (Shukla et al., 2016).

We exploit the feature extraction capabilities of small networks of VO₂ coupled oscillator to replace digital filters in CNNs **(figure 1 b)**. We fabricate VO₂ oscillators on a Si platform, adopting a crossbar

**Coupled VO$_2$ oscillators circuit as analog first layer filter in convolutional neural networks**

(CB) configuration with scaled device dimensions down to 70 nm. The CB devices exhibit improved variability and reliability compared to co-planar structures and enable the coupling of 4 oscillators. We demonstrate that such a 4-node ONN can memorize and perform 5 different filtering actions of a CNNs in a single circuit. Simulations with a 3x3 ONN further show how the concept can be applied to replace digital filters in the first layer of a CNN with a VGG-13 inspired architecture and through the adoption of a transfer learning technique. The hybrid CNN-ONN platform has been tested on the MNIST algorithm reaching recognition performances up to 95%. As an outlook, we discuss the benchmark of this technology when extended to all the layers of a CNN, up to the fully connected layers, in comparison with existing hardware and conclude that ONN might be used as fast and low-power inference machines.

## 2    Method

### 2.1    Device fabrication

VO$_2$ is a phase change material that presents a volatile, temperature driven insulator to metal transition (IMT). The transition can be triggered by joule heating when a voltage is applied to a VO$_2$ device (Kim et al., 2010). VO$_2$ can be grown crystalline on TiO$_2$ substrates; however, when deposited on Si, the film forms grains of the average dimension of ~50 nm (Premkumar et al., 2012). In the interest of future integration with electronics we have chosen to focus on integration on silicon in our work.

**Figure 2** shows VO$_2$ devices fabricated in a CB geometry on a 4'' Si wafer with a 1 μm thermal SiO$_2$ layer. Trenches are etched into the SiO$_2$ substrate and filled with Pt to provide the bottom electrode. Subsequently, an 80 nm thick VO$_2$ film is grown via atomic layer deposition and post-annealed, resulting in a policrystalline, granular film. Finally, top electrodes are formed using e-beam lithography and Pt evaporation. The smallest device area is 70 nm x 70 nm allowing a very compact design. The resistivity versus temperature curve (RT) of a 250 x 250 nm device is shown in **figure 3 a** and exhibits an insulator-to-metal phase-transition with roughly a two-order of magnitude in resistance change. The step-like RT implies multi-grain transitions, as already shown in previous work (Ruzmetov et al., 2009; Takami et al., 2014; Corti et al., 2019). **Figure 3 b** shows the insulator-to-metal and metal-to-insulator transition of an electrically activated device. A current source is used to control the current flowing in the device; a voltmeter is used to measure the voltage at each point. The IV characteristic of this device shows three different operating regions: a first region, in which the device is in its high resistance state, a negative differential resistance regime upon the phase change, and finally the low resistance region. Compared to previous work with planar VO$_2$ structure (Corti et al., 2019), we report that the crossbar structure allows for a decrease in voltage threshold device-to-device variability from 20% to 10% and the resistance-to-resistance variability from 10% to ~5%. Compared to other demonstrations on silicon, the improved variability allows for coupling of more oscillator nodes, up to 4. However, to go to larger network, careful material and device development is necessary to bring this figure down. The devices are tested in temperature-controlled chamber at 320 K and connected in the circuit configurations through external electrical components.

### 2.2    Oscillatory Neural Network

A single oscillator unit is realized biasing the VO$_2$ device in the negative differential resistance regime with a series transistor as described in (Parihar et al., 2014). Biasing the device in its negative differential resistance regime, the VO$_2$ switches continuously between its insulating and metallic state, originating relaxation oscillations at the drain voltage of the transistor. The oscillators are coupled via resistive and capacitive elements, as shown in **figure 4**, which ensure frequency and phase-locking of the drain voltage signals. The strength of the coupling element $C_{ij}$ that connects oscillator $i$ with oscillator $j$ can be calculated starting from the patterns to be memorized, via the Hebbian Learning Rule (HRL) (Hoppensteadt and Izhikevich, 1999):

$$C_{ij} = \frac{1}{n} \sum_{k=0}^{m} \vartheta_i^k \overline{\vartheta_j^k}$$

where $n$ is the total number of pixels per each image, or equivalently the number of oscillators in the network, $\vartheta_i^k$ is the value associated to the pixel $i$ of pattern $k$, and $m$ is the total number of patterns to be memorized in the ONN **(figure 5)**. These values $C_{ij}$ are then translated in different values of the coupling resistance $R_c$ between the oscillators. The memorized patterns appear in the operating ONN as stable phase relations between each oscillator $i$ and $j$. An oscillator in phase with the reference oscillator is translated into a white pixel; an oscillator with 180° phase difference with the reference corresponds to a black pixel. Given $m$ patterns memorized in the oscillatory neural network, the oscillators can stabilize their phase only according to one of the $m$ memorized patterns. When the oscillators are initialized to an unstable phase relation, they will relax to the nearest stable ensemble of phase relations, i.e. to the nearest memorized pattern. In this way, from a distorted pattern a memorized pattern is retrieved. In our system in **figure 4**, the oscillators are initialized to have different phase relations via a delay in the switching voltages of the single oscillators. The output is represented by the phase of the oscillating transistor drain voltage, compared to a reference. The phase stabilizes after around 5 oscillation periods for ONNs like shown in figure 4. The success of the input time-delay technique for image recognition is explained in detail in (Corti et al., 2020). In the experiments presented in this paper, the input delay signal is generated through a signal generator unit (National Instruments), the oscillators' output is acquired by a signal acquisition set-up and the phase calculated with post processing. The circuit coupling elements are realized with external electrical resistances. As an outlook, the input time-delay can be implemented in hardware via ring oscillators, and the phase-to-digital conversion can be tackled as described in (Staszewski et al., 2006). Also, the coupling resistance can be implemented with reconfigurable phase change memories (Boybat et al., 2018).

The simulations for the ONNs have been done using a Spice simulator. The VO₂ device was simulated with a behavioral model as described in (Maffezzoni et al., 2015). TensorFlow™ was used for the CNN and the hybrid ONN-CNN algorithms.

## 3 Results

### 3.1 Four-coupled oscillators

In this section we present a demonstration of four coupled VO₂ oscillators on Si, in which multiple patterns can be memorized. To form relaxation oscillator circuits, the VO₂ resistors on a silicon wafer are coupled through externally connected resistors and capacitances. An example of the measured waveforms of 4 coupled oscillators is shown in **figure 6**. The oscillators appear to be locked in frequency and the phase relation is calculated taking the distance between the crossing of the 1 V line in the falling edge of the oscillator curves. The coupling network has been programmed to recognize features as in a first layer of a convolutional neural networks. Looking at available analysis of feature extraction in convolutional neural networks (Zeiler and Fergus), the filters in the first layer commonly select edge features, like borders, diagonal, horizontal and vertical edges. Therefore, for the experimental demonstration, the ONN was trained to store vertical, horizontal and diagonal patterns. The weights of the circuit elements were identified through the Hebbian learning rule. To the best of our knowledge, this is the first demonstration of 4 coupled VO₂ oscillators with memory capabilities realized on a silicon platform. The circuit parameters used for the experiments are: $R_{12}$, $R_{13}$, $R_{24}$, $R_{34}$ = 82 k$\Omega$, $R_{23}$, $R_{14}$ = 130 k$\Omega$, $C_c$ = 5,6 nF, $V_{gx}$ = 1,4-1,6 V, $V_{in}$= 1.8-2.2 V. The different values of gate voltages $V_g$ and of the input signal $V_{in}$ are used to achieve similar frequency for the oscillators, that presents around 10 % of device-to-device variability. The horizontal, vertical and diagonal patterns are identified over multiple experiments, as depicted in **figure 7**. In addition, a fourth pattern in which all the oscillators result equally spaced was identified. The measurements are performed assuming Oscillator 1 as the reference oscillator; the phase of the other oscillators is calculated in respect to the crossing of the 1 V threshold of oscillator 1. Therefore, oscillator 1 has always a phase equal to 0, with a minimal data scattering that is calculated taking into account the variability of the value of the first experimental point that crosses the 1 V line. The other oscillators present a larger scatter, which doesn't impair the clear identification of the various patterns. However, random fluctuation of the oscillations and cross-talk noise hindered the experimental pattern recognition using the input-delay to output phase inference process. This is expected to improve with further process and design optimization of the crossbar devices.

To demonstrate the filtering capabilities of the circuit on an entire image, without suffering from the non-idealities of the experimental demonstration, circuit simulations calibrated on the experiments were conducted using Spice, reducing the variability of the VO₂ devices was from 10% to 5% and therefore increasing the recognition accuracy. In these simulations all the 4 patterns identified in the experiments were also observed; in addition, when the input delays of the circuit were chosen to be all the same, the oscillators in simulations were all oscillating in the in-phase configuration. This is an example of identification of a spurious pattern that was not encoded with the HRL. Spurious patterns arise when the memory capacity of the oscillatory neural network, that is studied to be $0.15n$ patterns for a n-oscillator network, is violated (Takashi Nishikawa, Ying-Cheng Lai, 2004; Follmann et al., 2015). Nevertheless, in such small oscillator networks the spurious patterns can be harvested as additional information. As shown in **figure 8,** when using the 2x2 ONN filter on an image of the MNIST dataset, vertical, horizontal and diagonal edges can be identified. In addition, the background as well as the images parts that have little contrast, can be identified through the in-phase oscillating condition. This demonstrates that a single ONN filter can operate as convolutional feature edge extraction identifying 5 different features.

### 3.2 ONN-CNN

Having shown that our simulations can reproduce experimental behaviour, we extend the simulations to explore the use of ONNs in combination with CNNs. The ONN circuit described in section 2.2 is simulated with Spice simulations using for the VO₂ device a behavioral model as described in (Maffezzoni et al., 2015). The simulations are done with both 2x2 and 3x3 oscillators ONNs based on parameters extracted from experimental devices. A convolutional neural network with a structure similar to a VGG-13 is trained on the MNIST dataset with a standard back-propagation algorithm **(Table 1)**. The trained weights are used to identify which features are recognized in the first layer of the CNN, that comprises 64 filters with a dimension of 3x3. In our network, as in (Zeiler and Fergus), it was also possible to identify multiple filters that selected horizontal, vertical and diagonal edges. We use the Hebbian Learning rule to store the same patterns in a 3x3 ONN matrix. The matrix dimension was chosen according to the dimension of the first layer convolution matrixes in the CNN. 10000 images from the MNIST dataset have been processed by the ONN matrix with a stride of 2, recognizing in each image vertical, diagonal, horizontal edges and uniform background. As already mentioned, storing of more than $0.15\ n$ patterns, where n is the number of the oscillators (Follmann et al., 2015), results in the appearance of spurious patterns that can in principle hinder the feature edge extraction process. However, As already discussed for the 4-coupled oscillators experiments, the arising of spurious patterns isn't detrimental for feature extraction operations. In the 3x3 filter case, we derived the pattern information from 3 key oscillators that oscillate in-phase for each memorized edge. For example, referring to what is depicted in **figure 9**, each time oscillators 2, 5 and 8 oscillate in-phase a vertical edge is recognized, and similarly for the other edges.

With this technique, a dataset of 10000 images filtered by the single ONN was calculated, with dimensions 13x13x5, where 5 represent the number of features recognized by the single ONN filter. The dataset was split in 6000 training images and 4000 test images.

Subsequently, five filters in the pre-trained CNN that provide the same filtered images were identified and replaced by the ONN with a simple transfer learning process:

- The 64 CNN filters were convolved with the same images from the MNIST dataset and activated with a Relu function.
- The CNN-filtered images were compared to the ONN-filtered images calculating the mean square error; the minimum of the mean square error was used to identify the filters from the CNN that can be substituted with the ONN.
- A new dataset is created after the first layer, substituting the images filtered by 5 CNN filters with the 5 filtered images from the ONN.

The remaining neural network layers are trained on the new dataset, achieving a recognition accuracy on the training set of 100% and on the test set of 95%. The original CNN, in comparison, reported better accuracy on the test set, of 97%. The reason for the worsening of the neural network performances is attributed to the cases in which the ONN fails the feature edge extraction. In fact, insufficient training of the ONN (just using HLR) also leads to recognition errors. The implementation of a backpropagation algorithm to the ONN layer would allow to increase the recognition performance in the network. We therefore implemented and tested a backpropagation scheme in our simulations. In **figure 10** we show an input image feature that should be recognized as a vertical edge. However, when the ONN is trained with the HLR the recognition fails. A cost function $C = (\varphi_{train} - \varphi_{out})^2/2$ is calculated from on the phases of the desired output $\varphi_{train}$ and the obtained output $\varphi_{out}$. Assuming an exponential dependence of the rising and falling edge of the relaxation oscillator waveforms, the derivative in time can be derived and an improved coupling matrix calculated. During subsequent epochs of this training the phase error is reduced. In the example shown in figure 10, the feature is recognized after 8 epochs of training. While blurred features (allowing 40% grey scale) were only recognized with 30% probability using the untrained ONN, 100% of the features were recognized with the trained ONN. The extension of the backpropagation algorithm to the entire ONN-CNN is yet to be implemented, but is expected to boost the recognition performance. In addition, the direct implementation of the backpropagation algorithm would allow for direct training of a CNN algorithm on an ONN platform and should ultimately result in an increase of the training speed. In fact, 45 parameters need to be trained for 5 CNN filters of 3x3 pixels size, however only 36 parameters need to be trained for a single ONN that performs all filtering actions. The number of parameters to be trained is therefore reduced of 20%: this can represent an important advantage in terms of speed and power consumption when training larger networks.

## 3.3    Benchmark

In this section we benchmark the convolution operations conducted with the ONN compared to a conventional CPU or GPU. We assume that the first layer of the convolutional neural network presented in this paper is integrally realized via ONN filters operating in parallel. The first layer of the CNN consists of 64 filters of 3x3 dimension passing through a 27x27 pixel image with a stride of 2, accounting to total of 13x13 operations per filter. Assuming that each ONN can perform 5 filtering actions inherently, a total amount of $13 \times 13 \times 64/5 \approx 2200$ ONNs is required, which corresponds roughly to 20`000 oscillator units and 80`000 memristors for implementing the coupling. Assuming a minimum feature size of 100 nm for both the VO₂ oscillators as well as the memristor, the total estimated area would be around 0.001 mm².

For calculating the power consumption of the circuit, we refer to (Shukla et al., 2016; Corti et al., 2020), that demonstrate operations of the oscillators at the power $P = 20\ \mu W$ with a scaled supply voltage <1 V and f = 3 MHz frequency operation. The total energy for the ONN to process one image with 64 filters at 3 MHz, including the waiting time of 5 oscillating period for the output stabilization, is calculated as

$$ P \times f \times 5 = 0.6\ \mu J/frame $$

Similarly, assuming the mean value of the coupling resistance to be around 100 kΩ, and the voltage drop across it 0.7 V, the total energy consumption of the memristors is calculated to be 3.4 µJ/frame.

Scaling of the device dimensions, it is envisioned that the VO₂ oscillator could be driven with 1 µW @ 0.3 V at a moderately increased oscillation frequency of 20 MHz. Moreover, through improved processing and resulting device uniformity, the coupling strength could be weakened allowing 1 MΩ coupling resistance (Shukla et al., 2015). The figure of merit for such a scaled system would improve by 3 orders of magnitude resulting in an energy consumption of 3 nJ/frame.

For conducting the same operation, a standard GPU needs to perform (13x13) convolutions x 64 filters x (3x3) pixels/filter = 97,344 multiply-accumulation operation, that correspond to around 200,000 flops. In Intel's CPU Core I9, which runs 1 TFLOP/s at 95 W, the total energy accounts for 20 µJ / frame; in the NVIDIA Tesla V100 GPU, that operates 120 TFLOP/s @ 300W, the total energy is 500 nJ / frame **(Table 2).** We can conclude that the ONN system, when built with the current VO₂ technology, is operating now at less power consumption of a conventional CPU, and given the scaling capabilities presented in other works, has the possibility of outperforming the top GPU available on the market. This analysis has been conducted not considering the peripheral circuitry that the ONN system will require, and therefore should be taken just as a projection of the potentiality of this technology.

## 4    Conclusion

A concept for exploiting oscillatory neural networks as hardware accelerators in convolutional neural networks is presented in this paper. A 4-nodes oscillatory neural network was built with scaled VO₂ oscillators' technology on a Si platform. We show that the time-encoded output signal can store up to 5 trained filters and performs the equivalent function of multiple digital convolutional filters in a neural network. We expand the concept to a 3x3 VO₂-ONN trained with Hebbian learning rule and simulate back-propagation for performance optimization. With the 3x3 filter and a transfer learning approach, we show that multiple digital filters of a CNN can be trained on a single ONN platform, achieving competitive recognition performances.

## 5    Data Availability Statement

The datasets generated for this study are available on request to the corresponding author.

## 6    Author Contributions

EC and SK ideated and designed the concepts and experiments proposed in this work. EC was responsible for the device fabrication, the measurements, circuit simulations and the neural network coding; JJ conducted the characterization on the crossbar devices; KN and JR conducted the deposition and annealing of the VO$_2$ films; KM, BG and AI were involved in the discussion, experiment design, and editing of the manuscript and provided valuable input at multiple stages of this work. All authors contributed to the article and approved the submitted version.

## 7    Acknowledgments

We would like to acknowledge the support of our partners and the founding of the HORIZON 2020 NEURONN project (Grant no. 871501) and PHASE-CHANGE SWITCH Project (Grant no. 737109)

## 8    Conflict of interest

EC, JJ, KM, BG and SK are employed by IBM.

**Coupled VO₂ oscillators circuit as analog first layer filter in convolutional neural networks**

**Coupled VO₂ oscillators circuit as analog first layer filter in convolutional neural networks**

Images:

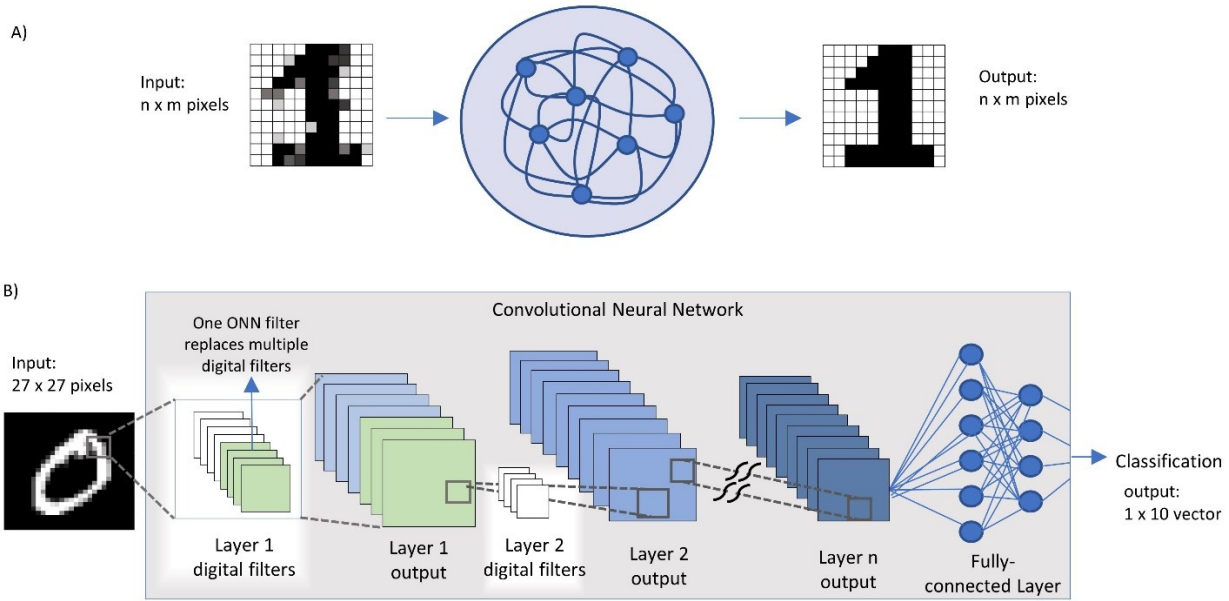

**Figure 1:** A) image inference process in an oscillatory neural network; B) image classification through a convolutional neural network

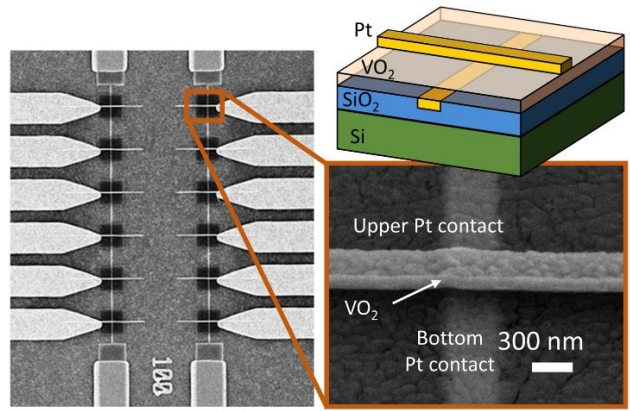

**Figure 2:** Left: scanning electron microscopy (SEM) image of 12 VO₂ devices. Top right: schematic of a VO₂ device deposited on a Si/SiO₂ substrate. The device area is defined by the width of the Pt contact lines. On the bottom, a SEM image of a 350 nm device. Minimum device dimension demonstrated: 70 nm.

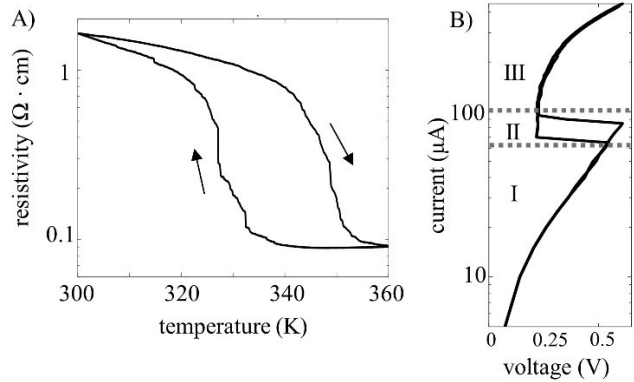

**Figure 3** A: resistivity measure of a 300x300 nm crossbar VO$_2$ device. The insulator to metal phase transition happens at around 340K and registers 2 orders of magnitude phase change. B: IV curve of a VO$_2$ device. Three different areas can be identified: the insulating region (I), the negative differential resistance region (II) and the metallic region (III)

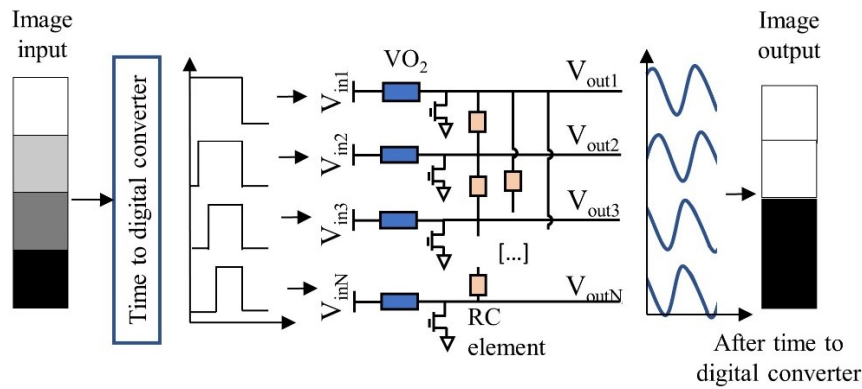

**Figure 4:** The coupled oscillator network serves as an image filter. The image input is converted into a delay of the oscillators' input signal. A single oscillator unit comprises a VO$_2$ phase-change element in series with a transistor. The coupling is realized with an externally connected resistance and a capacitance. The capacitance value is fixed, while the resistance value can be changed to store different patterns in the network and can be later substituted with a memristor.

**Coupled VO$_2$ oscillators circuit as analog first layer filter in convolutional neural networks**

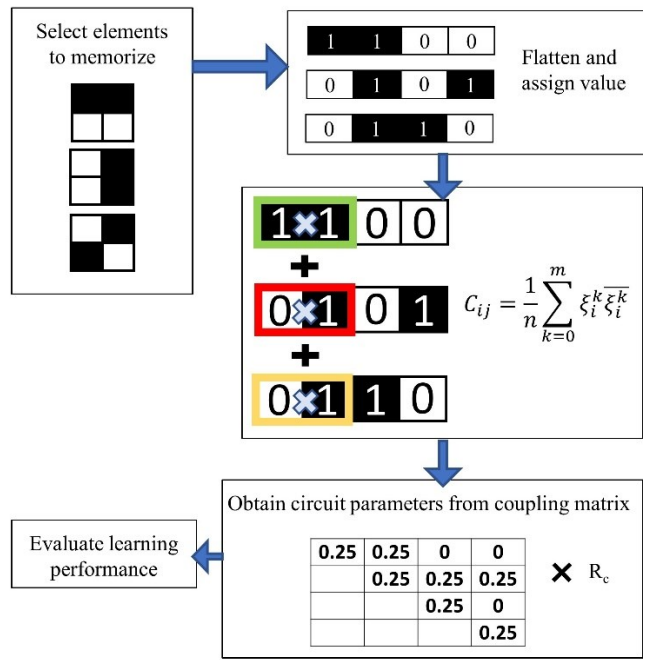

**Figure 5:** Flow chart of the learning. Weights are assigned to the pixel of each training image; the Hebbian Learning Rule is used to compute the coupling weights, which are translated into circuit values of the coupling resistance R$_c$.

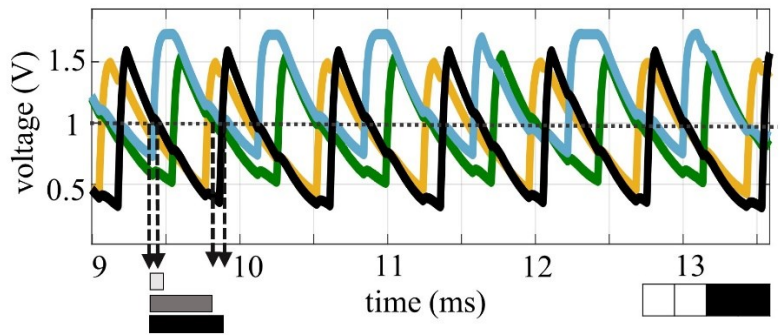

**Figure 6:** coupling of four VO$_2$ on Si oscillators. For reliable coupling, a hybrid R-C scheme was used, and the relative phase is calculated when the falling edge of the oscillations cross a 1 V threshold. In this experiment, an external capacitance of 150 nF was used on purpose to slow the oscillations, to enable a more precise sampling of the output signal.

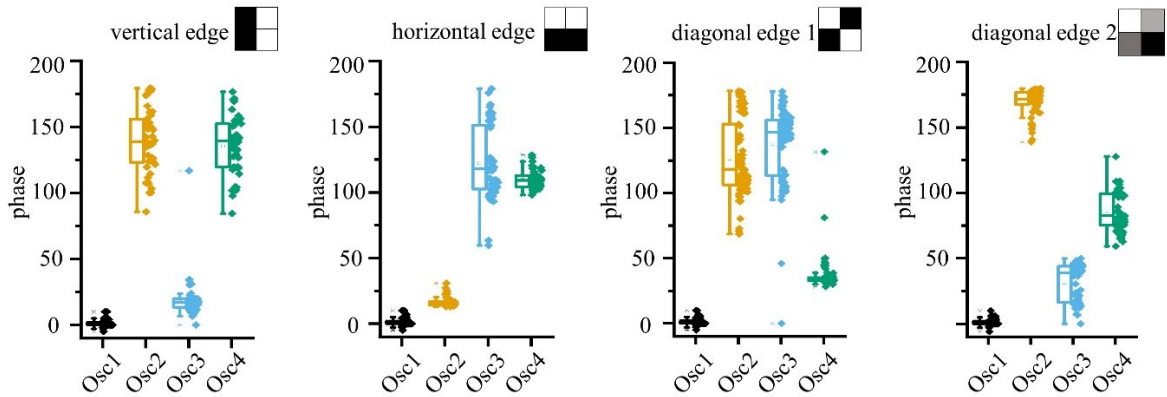

**Figure 7:** Experimental phase data that demonstrate that 4 features (e.g. vertical, horizontal and 2 diagonals) can be stored simultaneously in one 4-oscillators network. The pattern can be controlled by the time-delay of the oscillator drive. Phase-noise due to device variability impedes practical application. Circuit parameters: $R_{12}$, $R_{13}$, $R_{24}$, $R_{34}$ = 82 k$\Omega$, $R_{23}$, $R_{14}$ = 130 k$\Omega$, $C_c$ = 5,6 nF, $V_{gx}$ = 1,4 - 1,6 V, $V_{in}$= 1.8-2.2 V

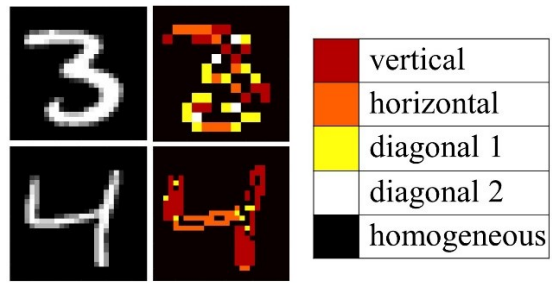

**Figure 8:** Simulation of convolution operation on MNIST images with a 2x2 VO₂ oscillator filter, which corresponds to 5 digital filters of the first layer of a CNN. The simulations are calibrated with the experimental results in figure 7.

**Coupled VO₂ oscillators circuit as analog first layer filter in convolutional neural networks**

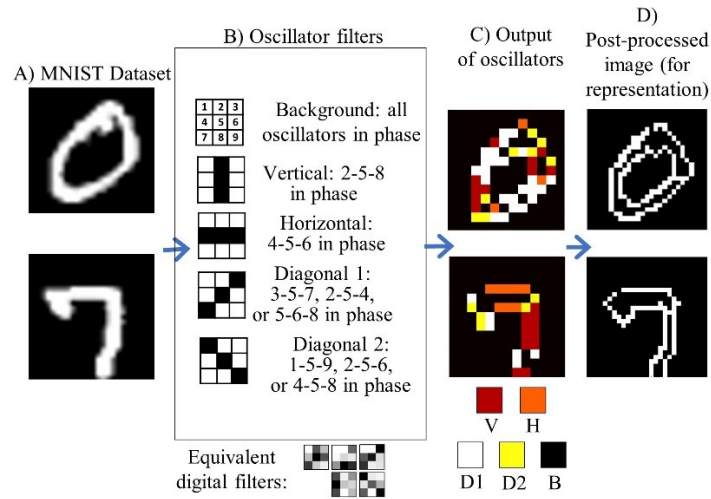

**Figure 9:** Extension of the convolution filtering operation to a 3x3 oscillator matrix with stride of 2. Five digital filters are replaced with 1 ONN filter that performs equivalent actions. The 9-bit information output is compressed in a 5-bit fashion for better representation of the edge direction in figure C. In figure D, the image in C is post-processed and expanded to a 27x27 pixel image to show the effectiveness of the ONN filter in recognizing the image features.

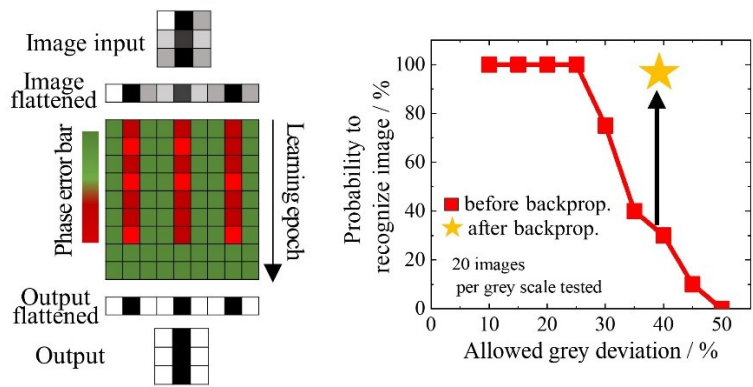

**Figure 10:** Example of backpropagation algorithm applied to a vertical edge recognition problem of the VO₂ ONN filter. Left: A distorted edge is given as an input to the ONN filter. The difference between the expected output phase and the output phase of the filter is depicted as pixel coloring from green (no phase error) to red (phase error). At the beginning the ONN filter fails the recognition, but after 8 learning epochs the filter is able to recognize the edge as a vertical edge. Right: the backpropagation algorithm allows the recognition of the image for increasingly distorted input features. This backpropagation algorithm is suitable for implementing filter training in an ONN-CNN.

**Coupled VO₂ oscillators circuit as analog first layer filter in convolutional neural networks**

| MNIST dataset | 27x27x1000 |
|---|---|
| ONN-CNN<br>5 ONN filters + 59 CNN filters | 3x3x64<br>stride = 2, padding = same |
| CNN1 | 3x3x64<br>stride = 1, padding = same |
| Max Pool 1 | 2x2<br>stride = 2, padding= same |
| CNN 2(x2) | 3x3x128<br>stride =1, padding = same |
| Max Pool 2 | 2x2<br>stride = 2, padding= same |
| CNN 3(x2) | 3x3x256<br>stride =1, padding = same |
| Max Pool 3 | 2x2<br>stride = 2, padding= same |
| Fully connected 1 | 4096 |
| Fully connected 2 | 1000 |
| Fully connected 3 | 10 |

**Table 1:** Schematic of the convolutional neural network architecture used in this work for performing the MNIST classification task. The network architecture is inspired by the VGG-13 architecture.

| | ONN<br>(current) | ONN<br>(projected) | CPU<br>Intel's Core I9 | GPU<br>Tesla V100 |
|---|---|---|---|---|
| Frames/s | $0.6 \times 10^6$ | $20 \times 10^6$ | $5 \times 10^6$ | $600 \times 10^6$ |
| Energy/frame | 3.400 μJ | 3 nJ | 20 μJ | 500 nJ |
| TFLOP/s | 0.12 | 0.75 | 1 | 120 |
| TFLOP/s $W^{-1}$ | 0.06 | 67 | 0.01 | 0.4 |

**Table 2:** Benchmark of the ONN technology against currently available platforms for convolutional neural network applications